\documentclass[12pt]{scrartcl}
\usepackage[utf8]{inputenc}

\usepackage{amsmath}
\usepackage{amssymb}

\newcommand{\wn}{\mathop{\text{wn}}}
\newcommand{\diag}{\mathop{\text{diag}}}
\newcommand{\pf}{\mathop{\text{pf}}}
\newcommand{\half}{{\textstyle \frac{1}{2}}}

\usepackage{marginnote}

\title{Topology of time-reversal invariant energy bands with adiabatic structure}
\author{Omri Gat\thanks{Permanent address: Racah Institute of Physics, The Hebrew University, Jerusalem 91904, Israel}~~and JM Robbins\\School of Mathematics, University of Bristol\\ University Walk 
Bristol BS8 1TW, UK.}
\begin{document}
\maketitle
\begin{abstract}
We classify the topology of bands defined by the energy states of quantum systems with scale separation between slow and fast degrees of freedom, invariant under fermionic time reversal. Classical phase space transforms differently from momentum space under time reversal, and as a consequence the topology of adiabatic bands is different from that of Bloch bands. We show that bands defined over a two-dimensional phase space are classified by the Chern number, whose parity must be equal to the parity of the band rank. Even-rank bands are equivalently classified by the Kane-Mele index, an integer equal to one half the Chern number.
\end{abstract}
\section{Introduction}
Spectral bands arise in stationary quantum systems in two principal ways. When a system has a non-compact symmetry group, states are labelled by  continuous representation indices. A prominent example is a particle in a periodic potential, whose states are labelled by the lattice momentum. The other possibility is that there is a separation of time scales, where the slow degrees of freedom control adiabatically the dynamics of the fast degrees of freedom. The dynamics of molecules in which the nuclei are several of orders of magnitude heavier than the electrons is an example of this type. The limit where the scale separation tends to infinity can be obtained formally by letting the effective Planck constant tend to zero for the slow subsystem, and energy bands (or energy surfaces in the Born-Oppenheimer terminology) are parametrised by the slow positions and momenta, belonging to a classical phase space.
 
In both cases the spectral problem reduces to the diagonalisation of a band Hamiltonian $H$ depending parametrically on either the lattice momenta or the 
slow phase space coordinates, so that $H$-invariant subspaces of each set of bands separated by a gap from the rest of spectrum define a vector bundle over the space of parameters. Physical observables depending only on the topological properties of these band vector bundles are \emph{robust}, in the sense that they do not change under continuous changes of the Hamiltonian as long as the gaps do not close.  

The Chern number, which classifies vector bundles over two-dimensional manifolds \cite{Nakahara:2003wv}, has the physical significance of quantum Hall conductance in periodic potential systems \cite{Thouless:1982wi} and anomalous band multiplicity in systems with slow degrees of freedom \cite{PavlovVerevkin:1988ue,Faure:2000tw}. In periodic potential systems the Chern number is mapped to its negative under time reversal (TR), and therefore Bloch bands in time-reversal invariant (TRI) systems have zero Chern number. Nevertheless, Kane and Mele have shown \cite{Kane:2005gb} that when TR is fermionic, so that the TR operator squares to $-1$, there are TRI bands that can be trivialised only by relaxing  TRI. Such bands are characterised by a nonzero value of the $\mathbb{Z}_2$-valued KM index, observable experimentally through the presence of conducting edge states. This  idea has been generalised to systems with more than two dimensions, other discrete symmetries, and systems with disorder and interactions, giving rise to the field of topological insulators \cite{Hasan:2010ku,Qi:2011hb}.

In contrast, an analogous theory has not been developed for adiabatic systems. Iwai and Zhilinskii \cite{Iwai:2011ft,Iwai:2012dz,Iwai:2014cz} studied Chern numbers and multiplicity of bands in systems with discrete \emph{unitary} symmetries, but the role of TR in systems with slow-fast structure is still not well understood. In this paper we address the following fundamental questions regarding fermionic TRI in adiabatic systems:
\begin{enumerate}
\item What are the topological classes of TRI vector bundles defined over a classical phase space?
\item Is there a topological $\mathbb{Z}_2$ index associated with these bundles?
\end{enumerate}

We answer these questions for vector bundles of arbitrary rank over a compact two-dimensional phase space, showing that the classification of TRI adiabatic bands is completely different than that of periodic systems bands. The first hint for this difference arises from the simple observation that, since configuration-space coordinates are invariant under TR while momenta change sign, TR is orientation-reversing on  a two-dimensional phase space, while it is orientation preserving on two-dimensional  momentum space. This implies that the Berry curvature is \emph{equal} at time-reversed points of adiabatic bands, rather than being of opposite signs, so that the Chern number does not necessarily vanish. Another difference is that whereas the TRI points in momentum space are  isolated, there are either no TRI points or else a line of TRI points in phase space; when there are no TRI points, the bundle rank does not have to be even.

Our first example is the two-sphere phase space of a rigid rotor, where time-reversal is the antipodal map.  We show the following:
\begin{enumerate}
\item TRI bundles are classified by the Chern number, which has to be even for bundles of even rank, and odd for bundles of odd rank (section \ref{sec:antipodal}). It follows that TRI bundles that are equivalent in the generic sense are also equivalent in the TRI sense.
\item It is possible to define an analog of the KM index for even-rank bundles, but here it is an integer index, equal to half the Chern number (section \ref{sec:kmindex}).
\end{enumerate}

The second example is the two-torus phase space of a particle in a periodic one-dimensional lattice, on which TR acts by mirror reflection (section \ref{sec:fpline}). The results for the TRI torus are very similar to those for the TRI sphere, except that only even-rank bundles are allowed because of the presence of TRI phase space points, namely those with zero or maximal momentum.

\section{Bands on the phase-space of a rigid rotor}\label{sec:antipodal}
When the slow dynamics consists of rigid body rotations, it is natural to label the band Hamiltonian by points on the two-sphere $S^2$, since the classical state is specified by the three components of the angular momentum, and in the absence of external torques the total angular is conserved; the natural area form is then the symplectic form of the $S^2$ phase space \cite{Arnold:1989wt}. Since angular momentum is odd under time reversal, time reversal sends a point $p$ on $S^2$ to its antipode $-p$.

We assume that the Hilbert space of the fast degrees of freedom is of finite dimension $N_A$, where $N_A$ is even. The eigenvalues of the fermionic TRI band Hamiltonian define $N_A$ continuous functions on $S^2$---the energy bands. The set of  $H$-invariant subspaces  of the band Hamiltonian $H$ associated with $N_B$ energy bands separated from the rest of the spectrum everywhere on the sphere form a rank-$N_B$ \emph{TRI vector bundle} $E(p)$ on $S^2$, i.e.~a rank-$N_B$ subbundle of $S^2\times\mathbb{C}^{N_A}$ such that $TE(p)=E(-p)$, where $T$ is an anti-unitary operator on $\mathbb{C}^{N_A}$ with $T^2=-1$.
Two TRI bundles $E_{0}(p)$ and $E_1(p)$ are \emph{TRI-equivalent} if there exists a continuous $E(p,s)$ such that $E(p,0)=E_0$, $E(p,1)=E_1$, and $E(p,s)$ is a TRI bundle for each fixed $s$.

The main results of this section are:
\begin{enumerate}
\item The parity of the Chern number of a TRI bundle is equal to the parity of its rank.
\item Two bundles are TRI-equivalent if and only if their Chern numbers is equal.
\end{enumerate}
To show this, let the two-sphere be embedded in the usual way in $\mathbb{R}^3$, and name the subset of the two-sphere with nonnegative 3rd coordinate (say) as the northern hemisphere $\mathcal{N}$. $\mathcal{N}$ is contractible so we can choose a smooth orthonormal frame $u_n(p)\in E(p)$, $n=1,\ldots,N_B$, for $p$ in $\mathcal{N}$; $Tu_n(-p)$ is then an orthonormal frame in the southern hemisphere $\mathcal{S}=-\mathcal{N}$. On the equator $\mathcal{N}\cap\mathcal{S}$ the two frames are related by 
\begin{equation}\label{eq:uframeind}
Tu_n(\tfrac{\pi}{2},\phi+\pi)=\sum_{n'} U_{nn'}(\phi) u_{n'}(\tfrac{\pi}{2},\phi)
\end{equation}
for some unitary transition matrices $U(\phi)$, where $p = (\theta,\phi)$ denote standard spherical polar coordinates for $p\in S^2$. Viewed as column vectors, $u_n$ and $Tu_n$ define $N_A\times N_B$ frame matrices $\mathbf{u}$ and $T\mathbf{u}$ (respectively) in terms of which \eqref{eq:uframeind} takes the form 
\begin{equation}\label{eq:uframe}
T\mathbf{u}(\tfrac{\pi}{2},\phi+\pi)=\mathbf{u}(\tfrac{\pi}{2},\phi)U(\phi)^t.
\end{equation}

Apply time reversal to  both sides of (\ref{eq:uframe}) and replacing $\phi$ by $\phi+\pi$ gives
\begin{equation}
-\mathbf{u}(\tfrac{\pi}{2},\phi)=T\mathbf{u}(\tfrac{\pi}{2},\phi+\pi)U(\phi+\pi)^\dagger=\mathbf{u}(\tfrac{\pi}{2},\phi)U(\phi)^tU(\phi+\pi)^\dagger,
\end{equation}
which implies that
\begin{equation}\label{eq:uphipi}
U(\phi+\pi)^{t}=-U(\phi).
\end{equation}

The Chern number $c$ of the bundle is the winding number of $\det U(\phi)$ \cite{Nakahara:2003wv}. Equation (\ref{eq:uphipi}) implies that $\det U(\phi+\pi)=(-)^{N_B}\det U(\phi)$, from which it follows that the parity of $c$ is equal to that of $N_B$.

We now proceed in two steps. In subsection \ref{sec:nf}, we show that a gauge change, that is a smooth $p$-dependent change of frame in the fibres $E(p)$, brings $U(\phi)$ to a normal form (equation \ref{eq:nf} below) depending only on $c$.  Then, in subsection \ref{sec:homotopy}, we show that two bundles are TRI-equivalent if and only if they have the same normal form.

\subsection{Normal form frames}\label{sec:nf}
Let $U(N_B)$ denote the group of $N_B\times N_B$ unitary matrices.  A gauge transformation from $\mathbf{u}(p)$ to $\mathbf{v}(p)$, a new othonormal frame for $E(p)$, is described by $W(p) \in U(N_B)$ as follows: 
\begin{equation}\label{eq:gt}
\mathbf{v}(p)=\mathbf{u}(p)\overline{W(p)}.
\end{equation}
Since
\begin{equation}\label{eq:gt2}
T\mathbf{v}(\tfrac{\pi}{2},\phi+\pi)=\mathbf{u}(\tfrac{\pi}{2},\phi)U(\phi)^tW(\tfrac{\pi}{2},\phi+\pi)=\mathbf{v}(\tfrac{\pi}{2},\phi)W(\tfrac{\pi}{2},\phi)^tU(\phi)^tW(\tfrac{\pi}{2},\phi+\pi),
\end{equation}
the transition matrices $V(\phi)$ for the $v$-frame are related to  $U(\phi)$, the transition matrices  for the $u$-frame,  by
\begin{equation}\label{eq:gauge-tr}
V(\phi)=W(\tfrac{\pi}{2},\phi+\pi)^tU(\phi)W(\tfrac{\pi}{2},\phi).
\end{equation}

Our goal is to find a gauge which transforms  $U(\phi)$ to a specific $V(\phi)$; we therefore view (\ref{eq:gauge-tr}) as an equation $W(\tfrac{\pi}{2}, \phi)$, with $U(\phi)$ and $V(\phi)$ given. This equation can  be solved at $\phi=0$ by taking  $W(\tfrac{\pi}{2},0)=V(0)U(0)^{-1}$.  At $\phi =\pi$, we take $W(\tfrac{\pi}{2},\pi)=I$.  We then choose $W(\tfrac{\pi}{2},\phi)$ on $0 < \phi < \pi$ to be a continuous but otherwise arbitrary path in $U(N_B$)  from $W(\tfrac{\pi}{2},0)$ to $W(\tfrac{\pi}{2},\pi)$    (such a path exists, as $U(N_B)$ is connected).   For   $\pi < \phi < 2\pi$, $W(\tfrac{\pi}{2},\phi)$ is determined by (\ref{eq:gauge-tr}), which relates $W$ at $(\tfrac{\pi}{2},\phi)$ and $(\tfrac{\pi}{2},\phi+\pi)$, as follows:
\begin{equation}
 W(\tfrac{\pi}{2},\phi)=\bigl(U(\phi-\pi)^{-1}W(\tfrac{\pi}{2},\phi-\pi)^{-1}V(\phi-\pi)\bigr)^t, \ \ 
 \pi<\phi<2\pi.
\end{equation}

It remains to check that $W(\tfrac{\pi}{2},\phi)$ is continuous at $\phi=\pi$ and $\phi =2\pi$.  We have that
\begin{align}
W(\tfrac{\pi}{2},\pi+0)&= \bigl(U(0)^{-1}U(0)V(0)^{-1}V(0)\bigr)^t=1=W(\tfrac{\pi}{2},\pi),\\
W(\tfrac{\pi}{2},2\pi-0)&=\bigl(U(\pi)^{-1}V(\pi)\bigr)^t=\bigl((-U(0)^t)^{-1}(-V(0)^t)\bigr)^t=V(0)U(0)^{-1}=W(\tfrac{\pi}{2},0).\nonumber
\end{align}

Since the fundamental group of $U({N_B})$ is $\mathbb{Z}$, and the homotopy class of the loop $W(\tfrac{\pi}{2},\phi)$ in $U({N_B})$ is the same as that of $\det W(\tfrac{\pi}{2},\phi)$ in $U(1)$, the gauge transformation $W(\tfrac{\pi}{2},\phi)$ can be extended from the equator to the  northern hemisphere if and only if  $\det W(\tfrac{\pi}{2},\phi)$ has zero winding number, i.e.
\begin{equation}
\wn\det W(\tfrac{\pi}{2},\cdot)\equiv\int_0^{2\pi}\frac{d\phi}{2\pi}\partial_\phi\arg\det W(\tfrac{\pi}{2},\phi)=0.
\end{equation}
On the other hand, equation (\ref{eq:gauge-tr}) implies that 
\begin{equation}
\wn\det W(\tfrac{\pi}{2},\cdot) =\half(\wn\det V-\wn\det U),
\end{equation}
so that for any $V(\phi)$ for which  $\wn\det V=\wn\det U=c$,  a solution  $W(\tfrac{\pi}{2}, \phi)$ of equation (\ref{eq:gauge-tr}) can be extended to a gauge transformation $W(p)$ defined on the northern hemisphere. We can therefore find a gauge transformation for which the transition matrices are in normal form 
\begin{equation}\label{eq:nf}
V_\text{nf}(\phi) =\diag(e^{i(c-N_B+1)\phi},e^{i\phi},\ldots,e^{i\phi}),
\end{equation}
where $\diag(d_1,d_2,\ldots)$ is a shorthand for a (block)-diagonal matrix with $d_1,d_2,\ldots$ as diagonal entries.

\subsection{The Chern number and TRI equivalence}\label{sec:homotopy}
We first show that TRI-equivalent bundles have the same Chern number. Suppose that $E_0$ is TRI-equivalent to $E_1$. The TRI equivalence $E(p,s)$ is then a TRI bundle over $S^2\times[0,1]$ where time reversal sends $(p,s)$ to $(-p,s)$, and we can choose a frame $\mathbf{u}_s(p)$ for $E(p,s)$ over the contractible subset $\mathcal{N}\times[0,1]$, and define $s$-dependent transition matrices $U_s(\phi)$ using the $s$-dependent generalisation of (\ref{eq:uframe}). The winding number of $\det U_s(\cdot)$ is a continuous integer-valued function of $s$ and therefore is constant, so that the Chern numbers of $E_0$ and $E_1$ are the same.

To show the converse, let $E_{0}$ and $E_1$ be TRI bundles with a common Chern number $c$. Choose smooth normal form frames $\tilde v_{n,s}$ for $E_s$, $s=0,1$, with transition matrices $V_\text{nf}$, and extend them to TRI frames of the ambient space $\mathbb{C}^{N_A}$, such that the subframes $\tilde v_{n,s}$,
$N_B+1\le n\le N_A$ for the orthogonal-complement bundles $E_s^\perp$ are in normal form. The extended frame matrices $\tilde{\mathbf{v}}_{s}$ ($s=0,1$) define sections of the trivial $U({N_A})$ bundle over the northern hemisphere obeying 
\begin{align}\label{eq:ttrinf}
T\tilde{\mathbf{v}}_{s}(\tfrac{\pi}{2},\phi+\pi)&=\tilde{\mathbf{v}}_{s}(\tfrac{\pi}{2},\phi)\tilde V(\phi)\ ,\\ 
\tilde V(\phi)&=\diag(e^{i(c-N_B+1)\phi},e^{i\phi},\ldots,e^{i\phi},e^{i(N_B-c-N_A+1)\phi},e^{i\phi},\ldots,e^{i\phi}).
\end{align} 
Next, we define the $SU(N_A)$ sections $\mathbf{v}_{s}=\diag(1/\det\tilde{\mathbf{v}}_{s},1,\ldots,1)\tilde{\mathbf{v}}_{s}$, which obey
\begin{align}\label{eq:trinf}
T\mathbf{v}_{s}(\phi+\pi)&= \mathbf{v}_{s}(\phi)V(\phi)\ ,\\ 
V(\phi)&=\diag(e^{i\alpha+ i(c-N_B+1)\phi},e^{i\phi},\ldots,e^{i\phi},e^{i(N_B-c-N_A+1)\phi},e^{i\phi},\ldots,e^{i\phi})
\end{align}
where $\alpha$ is such that, for any $N_A\times N_A$ matrix $A$, we have that $\det(TA)=e^{i\alpha}\overline{\det A}$.

We say that a section $\mathbf{u}(p,s)$ of  the trivial $U(N_A)$-bundle over the cylinder $\mathcal{N}\times[0,1]$ is $N_B$-TRI if $\mathbf{u}_s(\frac{\pi}{2},\phi)^{-1}T\mathbf{u}_s(\frac{\pi}{2},\phi+\pi)$ is block diagonal with blocks of size $N_B$ and $N_A - N_B$. 
Then $E_0$ and $E_1$ are TRI-equivalent if there exists an $N_B$-TRI section $\mathbf{u}_s(p)$ of the trivial $U(N_A)$-bundle $\mathcal{N}\times[0,1] \times U(N_A)$,  such that the first $N_B$ columns of $\mathbf{u}_0(p),\mathbf{u}_1(p)$ span $E_0(p)$, $E_1(p)$ respectively.

We will complete the argument by constructing a frame $\mathbf{v}_s(p)$ that obeys \eqref{eq:trinf} for every $s$ between 0 and 1 (and is therefore $N_B$-TRI) and reduces to $\mathbf{v}_{0,1}(p)$ for $s=0,1$ (respectively). For this purpose, let 
$\mathring{\mathbf{v}}_s(n)$ be a continuous path in $SU(N_A)$ connecting $\mathbf{v}_0(n)$ and $\mathbf{v}_1(n)$, $n$ being the north pole, and define
\begin{equation}
\mathring{\mathbf{v}}_s(\theta,\phi)=\begin{cases} \mathbf{v}_0\bigl((1-2s)\theta,\phi\bigr)\mathbf{v}_0(n)^{-1}\mathring{\mathbf{v}}_s(n)&0\le s\le\frac12\\
\mathbf{v}_1\bigl((2s-1)\theta,\phi\bigr)\mathbf{v}_1(n)^{-1}\mathring{\mathbf{v}}_s(n)&\frac12\le s\le1\end{cases}
\end{equation}
for $0<\theta\le\frac{\pi}{2}$.

We now seek a gauge transformation ${\mathbf{v}_s}(p)= \mathring{\mathbf{v}}_s(p)\overline{W_s(p)}$, $W\in SU(N_A)$ satisfying (compare with (\ref{eq:gt2},\ref{eq:gauge-tr}))
\begin{equation}\label{eq:gauge-trcyl}
V(\phi)=W_s(\tfrac{\pi}{2},\phi)^t\mathring{\mathbf{v}}_s(\tfrac{\pi}{2},\phi)^{-1}T\mathring{\mathbf{v}}_s(\tfrac{\pi}{2},\phi+\pi)W_s(\tfrac{\pi}{2}, \phi+\pi)
\end{equation}
on $\partial\mathcal{N}\times[0,1]$. $W$ can be constructed by first choosing
\begin{equation}
W_0(\tfrac{\pi}{2},\phi)=W_1(\tfrac{\pi}{2},\phi)=W_s(\tfrac{\pi}{2},0)=1\ ,\ \  W_s(\tfrac{\pi}{2},\pi)=(T\mathring{\mathbf{v}}_s(\tfrac{\pi}{2},\pi))^{-1}\mathring{\mathbf{v}}_s(\tfrac{\pi}{2},0)^{-1}V_s(0).
\end{equation}

Since  $\det W_0(\tfrac{\pi}{2},\phi)=\det W_1(\tfrac{\pi}{2},\phi) = \det W_s(\tfrac{\pi}{2},0)=1$, $W$ can first be extended from the closed curve  defined by $(\theta=\tfrac{\pi}{2},\phi=0,0\le s\le1),(\theta=\tfrac{\pi}{2},0\le\phi\le\pi,s=1),(\theta=\tfrac{\pi}{2},\phi=\pi,1\ge s\ge0),(\theta=\tfrac{\pi}{2},\pi\ge\phi\ge0,s=0)$  to the surface it bounds on $\partial\mathcal{N}\times[0,1]$, that is, $\theta=\tfrac{\pi}{2},0\le\phi\le\pi,0\le s\le1$. These values then define $W$ on the surface $\theta=\tfrac{\pi}{2},-\pi\le\phi\le0,0\le s\le1$, through (\ref{eq:gauge-trcyl}). Next let $W\equiv1$ on the faces $\mathcal{N}\times\{0\},\mathcal{N}\times\{1\}$, so that $W$ is defined continuously on all of the boundary of $\mathcal{N}\times[0,1]$ and obeys (\ref{eq:gauge-trcyl}). Finally $W$ can be continuously extended to the interior of $\mathcal{N}\times[0,1]$ because the second homotopy group of $SU({N_A})$ is trivial.

\section{The Kane-Mele index on the two-sphere}\label{sec:kmindex}
In the previous section we showed that phase-space TRI vector bundles are characterised by the Chern number, which can be nonzero, unlike  momentum-space TRI where $c=0$, and bundles are classified by the Kane-Mele index $k$. Still, one may ask whether $k$ can be defined for phase-space bundles, and what topological information it conveys. We next show that $k$ makes sense as an \emph{integer} index on bundles of {even} rank on the two-sphere phase space, and that $k=c/2$.

There are several equivalent definitions of $k$ for bundles on momentum spaces. Of these, the one that is applicable to bundles on the two-sphere phase space is the original one given in \cite{Kane:2005gb}, which does not rely on the existence of TRI points in the base space. It is based instead on the $N_B \times N_B$ matrix $M(p)$ with elements $\langle u_n(p),T u_{n'}(p)\rangle$, where $u_n$ is a unitary frame for the bundle in the northern hemisphere; it follows from the definition that $M$ cannot be globally defined as a continuous function on the sphere when $c\ne0$, but this presents no difficulty in the analysis.

It also follows from the definition and $T^2=-1$ that $M$ is skew-symmetric, so we can define the Pfaffian $\pf M$. The Kane-Mele integer is then
\begin{equation}\label{eq:kdef}
k=\wn\pf M(\tfrac{\pi}{2},\cdot)
\end{equation}
We will only consider explicitly the generic case where the zeros of $\pf M$ are isolated, and then it always possible to choose the northern hemisphere such that $\pf M\ne0$ on the equator and the definition (\ref{eq:kdef}) of $k$ is valid.\footnote{Ref \cite{Kane:2005gb} has shown how to generalise the definition to the case where there are zero curves because of symmetry.}
$k$ then counts the total index of the zeros of $\pf M$ in the northern hemisphere. The positions and indices of these zeros do not change under a gauge transformation so that $k$ is gauge invariant. This also follows from the demonstration that $k=c/2$ which we present next (compare with appendix A of \cite{Fu:2006jk}.)

Let $U$ be the equator transition matrix of the frame $u$, defined as in (\ref{eq:uframeind}). Then 
\begin{align}
\langle u_n(\tfrac{\pi}{2},\phi+\pi),T u_{n'}(\tfrac{\pi}{2},\phi+\pi)\rangle&=-\langle T^2u_n(\tfrac{\pi}{2},\phi+\pi),T u_{n'}(\tfrac{\pi}{2},\phi+\pi)\rangle=\nonumber\\
&=-\sum_{n''n'''}U_{nn''}(\phi)\langle T u_{n''}(\tfrac{\pi}{2},\phi),u_{n'''}(\tfrac{\pi}{2},\phi)\rangle U_{n',n'''}(\phi),
\end{align}
so that 
\begin{equation}
M(\tfrac{\pi}{2},\phi+\pi)=-U(\phi)\overline{M(\tfrac{\pi}{2},\phi)}U(\phi)^t\Rightarrow \pf M(\tfrac{\pi}{2},\phi+\pi)=\pm \det U(\phi)\overline{\pf M(\tfrac{\pi}{2},\phi)},
\end{equation}
where the sign in the rightmost expression is positive or negative when the bundle rank is congruent to 0 or 2 modulo 4, respectively.

It now follows from (\ref{eq:uphipi}) that $\det U(\phi+\pi)=\det U(\phi)$ so that
\begin{equation}
k=\int_0^\pi\frac{d\phi}{2\pi i}\partial_\phi(\arg\pf M(\tfrac{\pi}{2},\phi)+\arg\pf M(\tfrac{\pi}{2},\phi+\pi))=\int_0^\pi\frac{d\phi}{2\pi i}\partial_\phi\arg\det U(\phi)=\frac{c}{2}.
\end{equation}

\section{Bands on the two-torus phase space}\label{sec:fpline}
Fermionic TR is usually associated with Kramers degeneracy, since TRI subspaces of $C^{N_A}$ must have even dimension. This is avoided in the rotor phase space of section \ref{sec:antipodal}, because the two-sphere has no TRI points. 
We now analyse the topology of bands of systems 
where the slow dynamics takes place in a one-dimensional lattice with periodic boundary conditions, so that the classical phase space is a two-torus, with the flat area form given by the symplectic form \cite{Hannay:1980wk,Hannay:1994ti}. Choosing canonical coordinates $q$ and conjugate momentum $p$ periodic modulo $2\pi$, so that that $q$ is even and $p$ is odd under time reversal, the TRI points belong to the two lines $p\equiv0,\pi\pmod{2\pi}$. Thus, unlike momentum spaces where the TRI points are isolated, here they are a part of lines of invariant points.

We show that, other than the presence of Kramers degeneracy, the topological properties of the two-torus phase space bands is the same as for the two-sphere phase space bands, so that
\begin{enumerate}
\item The rank and Chern number $c$ of a TRI band on the two-torus phase space must be even.
\item Two bands are TRI-equivalent if and only if they have the same Chern number.
\item The Kane-Mele index of the band $k=c/2$.
\end{enumerate}

Since the cylinder $0\le p\le\pi$ has a trivial second homotopy group \cite{Hatcher:2001ut}, we can  define on it a frame $\textbf{u}(q,p)$,
with transition matrices $U^\pm(q) \in U(N_B)$  given by
\begin{equation}\label{eq:uframe-t2}
T\textbf{u}(q,0)= \textbf{u}(q,0)U^+(q)^t\ ,\qquad
T\textbf{u}(q,\pi)=\textbf{u}(q,\pi)U^-(q)^t .
\end{equation}
The Chern number is then
\begin{equation}
c=\wn\det U^+-\wn\det U^- .
\end{equation}

The matrix $M_{nn'}(q,p)=\langle u_n(q,p),Tu_{n'}(q,p)\rangle$ is equal to $U_{nn'}^+(q)$ and $U_{nn'}^-(q)$ for $p=0,\pi$ (respectively), so that $U^\pm(q)$ are skew-symmetric. Since $\det M=(\pf M)^2$ it follows that
\begin{equation}
k=\wn\pf M(\cdot,0)-\wn\pf M(\cdot,\pi)=\frac{c}{2},
\end{equation}
establishing points 1 and 3 above.

Next we define a gauge transformation (\ref{eq:gt}) to a frame  $\textbf{v}(q,p)$,  with transition matrices $V^\pm(q) \in U(N_B)$ given by
\begin{equation}\label{eq:torusgauge}
V^+(q)=W(q,0)^tU^+(q)W(q,0)\ ,\qquad V^-(q)=W(q,\pi)^tU^-(q)W(q,\pi).
\end{equation}
The skew-symmetric unitary matrices $U^\pm(q)$ are unitarily congruent to block diagonal matrices of the form 
\begin{equation}\label{eq:uss}
V^\pm(q)=\diag\left(\begin{pmatrix}0&-e^{i\alpha_1^{\pm}(q)}\\e^{i\alpha_1^{\pm}(q)}&0\end{pmatrix}
,\ldots,\begin{pmatrix}0&-e^{i\alpha^{\pm}_{N_B/2}(q)}\\e^{i\alpha^{\pm}_{N_B/2}(q)}&0\end{pmatrix}\right)
\end{equation}
for any real $\alpha(q)_j^\pm$'s \cite{Horn:1622920}. If the $\alpha^\pm_j$'s are continuous modulo $2\pi$ in $q$, it is possible to choose the congruence matrices $W(0,q)$ and $W(\pi,q)$ continuously.

Let $\alpha_1^+(q)=\frac{c}{2}q$, and  take all other $\alpha_j$'s to be identically equal to $0$. 
Choose continuous functions $W(q,0),W(q,\pi)$ satisfying (\ref{eq:torusgauge}). Then
\begin{equation}
\wn\det V^+=c\ ,\qquad\wn\det V^-=0,
\end{equation}
and since
\begin{align}\nonumber
&\wn\det W(\cdot,0)=2(\wn\det V^+-\wn\det U^+),\\&\wn\det W(\cdot,\pi)=2(\wn\det V^--\wn\det U^-),
\end{align}
it follows that 
\begin{equation}
\wn\det W(\cdot,0)-\wn\det W(\cdot,\pi)=0,
\end{equation}
 and $W$ can be extended continuously to the entire cylinder $\mathcal{C}=\{(q,p)\, |\, 0\le q< 2\pi, 0\le p\le\pi\}$. The argument of section~\ref{sec:homotopy} adapted to $\mathcal{C}\times[0,1]$ shows that two bundles on the two-torus phase space are TRI-equivalent if and only if the have the same Chern number.

\section{Conclusions}
We studied the topological properties of vector bundles with fermionic TRI. Such bundles arise naturally in quantum mechanics of systems with discrete translational invariance and in systems with scale separation between slow and fast degrees of freedom. In the latter case, which is the focus of our work, the base is a classical phase space. 

The two-sphere studied in detail is the phase space of a rigid rotor, and is a good model for the rotation of a molecule, which is TRI in the absence of external magnetic fields.  TR is fermionic if the molecule has an odd number of electrons. A well-known example is the coupling of the total angular momentum to the spin of a single unpaired electron in molecules such as the hydroxyl radical, which
 leads to 
 $\lambda$-doubling \cite{Varberg:1993uk}. In this case, our results imply that the Chern number of the band is the only topological invariant and that it must be odd.

Similarly, the two-torus phase space is a model for trapped spin-polarised cold fermionic atoms in a one-dimensional optical lattice \cite{Giorgini:2008ct}. Here, Bloch bands are always Kramers-like degenerate and topology is classified by an even Chern number, or equivalently by the integer-valued Kane-Mele index, which  is equal to half the Chern number.

It turns out that the consequences of fermionic TRI in adiabatic systems   with two-dimensional slow phase spaces are markedly different than in the well-studied case of Bloch bands of fermions.  First, TRI does not preclude nonzero Chern numbers, but only restricts their parity.  Second,  two such TRI bundles equivalent in the generic sense (that is, via deformations that may violate TRI) are necessarily TRI-equivalent (that is, via deformations that preserve TRI).

The Kane-Mele index $k$ is a signed count, with sign determined by orientation, of the zeros of the Pfaffian of $M$, the matrix that relates a frame to its TR image, in a subset of the base space containing a single point from each TR-related pair.  It is well-defined in both the Bloch and adiabatic cases for bundles of even rank.  But whereas for Bloch bands, time-reversed zeros are oppositely oriented, they have the same orientation in a diabatic bands. It follows that the Chern number, which is the signed count  of zeros of $\pf M$ over the \emph{whole} base space, is always zero in the Bloch case, but in the adiabatic case it can be any even integer, and is equal to twice the Kane-Mele index.

Another difference is that $k$ is an integer invariant of adiabatic bands, while for Bloch bands it is only invariant modulo 2. $k$ can be changed for Bloch bands by $-2$, say,  by continuously transporting a positively oriented zero of $\pf M$ so that it annihilates the TR image of another positively oriented zero \cite{Kane:2005gb}.  Such an operation is impossible for adiabatic bands, because the time reverse of a positively oriented zero is also positively oriented.

Finally, it must be emphasised that the present results do not contradict the classification scheme of \cite{Schnyder:2008ez,Kitaev:bg,Ryu:2010ko}, according to which bundles with fermionic time reversal symmetry defined on a two-dimensional base are characterised by a $\mathbb{Z}_2$-invariant. Actually, the classification relies on the TR transformation behavior of base points \cite{Stone:2010bd}, which is different in Bloch bands and adiabatic bands. It is natural to ask what would be the analog of this classification for adiabatic bands, but the answer is beyond the scope of this work.

\textbf{Acknowledgements} We benefited from discussions with Ronnie Kosloff.
OG gratefully acknowledges support from the Leverhulme Trust Visiting Professorship program during his stay in Bristol.

\bibliographystyle{unsrt}
\bibliography{ti.bib}


\end{document}